# Production of Gadolinium-loaded Liquid Scintillator for the Daya Bay Reactor Neutrino Experiment

Wanda Beriguete[a], Jun Cao[b], Yayun Ding[b*], Sunej Hans[a], Karsten M. Heeger[c,d], Liangming Hu[a], Aizhong Huang[e], Kam-Biu Luk[f,g], Igor Nemchenok[h], Ming Qi[i], Richard Rosero[a], Hansheng Sun[b], Ruiguang Wang[b], Yifang Wang[b], Liangjian Wen[b], Yi Yang[b], Minfang Yeh[a**], Zhiyong Zhang[b], Li Zhou[b]

[a] Brookhaven National Laboratory, Upton, New York, USA

[b] Institute of High Energy Physics, Beijing, China

[c] University of Wisconsin, Madison, Wisconsin, USA

[d] Yale University, New Haven, CT, USA

[e] Jinling Petro-chemical Corporation, Nanjing, China

[f] Lawrence Berkeley National Laboratory, Berkeley, California, USA

[g] University of California, Berkeley, California, USA

[h] Joint Institute for Nuclear Research, Dubna, Moscow, Russia

[i] Nanjing University, Nanjing, China

**Abstract:** We report on the production and characterization of liquid scintillators for the detection of electron antineutrinos by the Daya Bay Reactor Neutrino Experiment. One hundred eighty-five tons of gadolinium-loaded (0.1% by mass) liquid scintillator (Gd-LS) and two hundred tons of unloaded liquid scintillator (LS) were successfully produced from a linear-alkylbenzene (LAB) solvent in six months. The scintillator properties, the production and purification systems, and the quality assurance and control (QA/QC) procedures are described.

*Corresponding author, tel: +86 10 88233215

**Corresponding author, tel: +1 631 3442870

Email addresses: dingyy@ihep.ac.cn (Y. Y. Ding), yeh@bnl.gov (M. Yeh)

**Introduction**

The Daya Bay reactor antineutrino experiment aims for a high-precision measurement of the neutrino mixing angle $\theta_{13}$ by searching for $\bar{\nu}_e$ deficiency as a function of distance from a nuclear reactor complex in Shenzhen, China [1-3]. Antineutrinos are detected via the inverse beta-decay (IBD) reaction, $\bar{\nu}_e + p \rightarrow e^+ + n$, in a delayed coincidence between the prompt signal of the positron and the subsequent capture of the neutron in a (n,γ) reaction after it has been thermalized in the scintillator. This delayed coincidence provides a unique $\bar{\nu}_e$ signature and serves as a powerful tool to reduce random backgrounds.

The neutron capture can occur on hydrogen in the liquid scintillator (LS). The cross section is 0.332 barns and the energy of the emitted gamma is 2.2 MeV. There are several important advantages of adding a neutron-capture-enhancing element, such as gadolinium (Gd), to the LS. The (n,γ) cross-section for natural Gd is 49,000 barns (with major contributions from the $^{155,157}$Gd isotopes), that only a small concentration of Gd, e.g. 0.1% by mass, is necessary. The neutron-capture on Gd releases 8-MeV of energy in a cascade of 3-4 γ-rays that can easily exclude the low-energy backgrounds from natural radiations in the surrounding environment. Furthermore, the neutron-capture time is significantly shortened to ~28 μs in a 0.1% Gd-LS, as compared to ~200 μs in LS. This shortened delay time reduces the accidental background rate by a factor of 7.

It is commonly known that the chemical stability of the Gd-LS is the key to the success of a reactor antineutrino experiment. Previous studies have identified that gadolinium could be either extracted or dissolved into a scintillator through chelating ligands, such as phorsphors, β-diketones, or carboxylates [4-9]. The Daya Bay collaboration uses linear alkylbenzene (LAB), a straight alkyl chain of 10-13 carbons attached to a benzene ring [10], as the detection medium due to its high flash point, low chemical reactivity, and light-yield comparable with other scintillators (e.g. pseudocumene). 3,5,5-trimethylhexanoic acid (TMHA) is chosen as the ligand because its Gd-complex has good stability in LAB and is easy to produce in large quantities [9]. The Daya Bay Gd-LS also consists of 3 g/L 2,5-diphenyloxazole (PPO) as the fluor and 15 mg/L p-bis-(o-methylstyryl)-benzene (bis-MSB) as the wavelength shifter. After years of development, a production scheme for the fabrication of a Gd-complex solid, followed by direct dissolution in scintillator was chosen to accommodate the logistics and operation at the Daya Bay nuclear power plant, and to produce stable Gd-LS for the Daya Bay experiment.

The Daya Bay detector system is composed of eight functionally identical antineutrino detectors (ADs) located in near and far underground halls close to the six reactor cores. Each AD has three liquid zones: gadolinium-loaded liquid scintillator (Gd-LS, neutrino target), unloaded liquid scintillator (LS, γ-catcher), and mineral oil (MO, shielding buffer) situated in the order of inner- to outer-most. Prior to the final production, two Gd-LS samples of 800 and 600 liters were prepared separately and deployed to the Institute of High Energy Physics (Beijing, China) [9] and the Aberdeen Tunnel Underground Laboratory (Hong Kong) [11] for prototyping tests. The performance of these Gd-LS liquids has been monitored for over a year since then and before the final production commenced to ensure that the chemical stability, optical transparency, and light-yield of the Gd-LS met the required criteria.

Major components of the production system for Gd-LS and LS were constructed and pre-tested in 2008. The purification of each raw material was studied, and the appropriate technologies and equipment were developed and constructed. The production scheme was optimized in various test runs. The actual manufacture of scintillator liquids started in late 2010 and finished in March 2011. A sum of ~400 tons of scintillator, including 185 tons of 0.1% Gd-LS and 200 tons of LS were produced in 4-ton batches over a period of six months. Both the Gd-LS and LS have been under long-term monitoring since then. The production, properties, quality assurance and control (QA/QC), and stability of these scintillators are presented in the following sections.

**1. Production System and Test Runs**

1.1 Production System

Figure 1(a) presents a three-dimensional diagram and 1(b) shows a photograph of the operation of the production system for synthesizing liquid scintillator, scaled up from the bench-top practices based on the compatibility of material choice and the safety requirement of underground confined space. The production system was first assembled for testing at the Institute of High Energy Physics (IHEP) and later disassembled for installation at the experimental site in 2010.

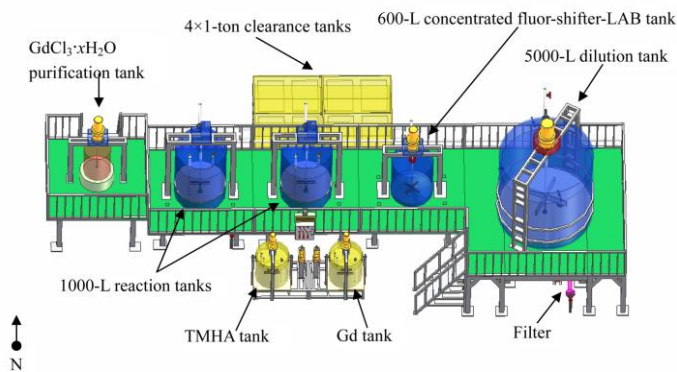
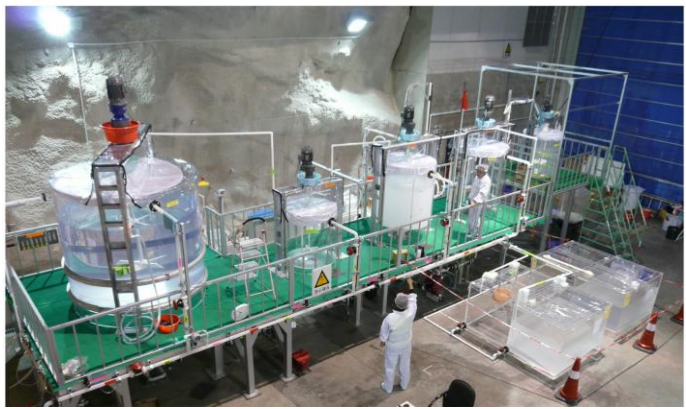

Figure1 (a) three-dimensional diagram (south to north) and (b) photograph of the operation of the production system for liquid scintillators (north to south).

Besides the production equipment, two 200-ton capacity concrete storage pools (A and B) and five 40-ton capacity acrylic tanks, together with the related piping system, were installed in the underground liquid scintillator hall (LS Hall) in 2009 to 2010. The two concrete pools were lined with thermoplastic Olefin (TPO) for secondary containment. Each pool contained a storage bag fabricated by the BLT Flexitank Industrial Co., Ltd. from nylon-66 and polyethylene composite membrane (PE/PA) produced by the Jiangyin Sunrise Packaging Material Co. Ltd. The inner side of the bag was chosen as nylon since polyethylene was found to contaminate LAB while nylon does not. These bags contained the LAB delivered from the vendor. The five 40-ton tanks for the storage of the synthesized Gd-LS were manufactured, cleaned and sealed in Gold Aqua Acrylic Co., Ltd. in Fujian, and then transported and installed in the LS Hall.

The manufacture of Gd-LS commenced in Oct. 2010. The LAB for Gd-LS production was taken only from storage pool-A. Following the completion of the Gd-LS production (pool-A was emptied) in Jan. 2011, the production of the LS started in Feb. 2011 and was completed in Mar. 2011. The LS produced from LAB taken from pool-B was transferred to pool-A for storage, and a new PE/PA bag was installed in pool-B for storage of mineral oil. Two months after the completion of liquid production, two ADs were filled and later commissioned side-by-side in the Daya Bay near hall (EH1) during the summer of 2011 [12].

1.2 Test Runs

Small batches of 0.1% Gd-LS were prepared under variable conditions in eight test runs at IHEP using the production equipment shown in Fig. 1 to establish the production procedure and optimize the synthesis parameters. The flow rates and agitation speeds were determined. Issues encountered in the scaling-up from the bench-top experiments were solved, such as the emulsification of the solution and the metal particle contamination. Additional shields were added to several compartments of the production system to prevent metal particles, i.e. iron generated by friction between the steel parts, from falling into the scintillator.

During the test runs, several key methodologies for measuring scintillator properties were also developed. The [Gd] concentration was determined either by back-extraction, followed by colorimetric titration using a UV-Vis spectrometer or by x-ray fluorescence (XRF). The [Gd] concentration of the Gd-LS produced from the several test runs averaged 0.093±0.003% from both methods. Absorbance spectra of the Gd-LS obtained with a UV-Vis spectrometer and a 10-cm quartz cuvette are shown in Figure 2. Differences between the absorption spectrum obtained immediately after production and that 1172 days later are insignificant. An absorbance at 430 nm of 0.003±0.001 corresponds to an attenuation length ($\lambda_{1/e}$) of 14±4 m. Attenuation lengths of scintillator components were further measured by a 1-m path-length system [13]. Values obtained for $\lambda_{1/e}$ at 430 nm were 10.5 m for Gd-LAB (without PPO and bis-MSB) and 14.1 m for pure LAB, which means that reasonable optical transparency of Gd-LS can be achieved even using unpurified starting materials of $GdCl_3$, TMHA, PPO, and bis-MSB.

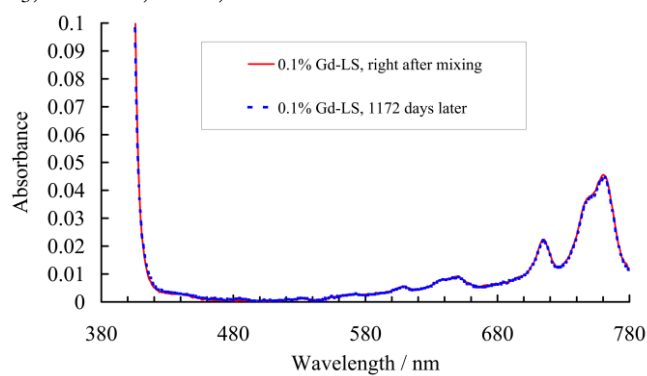

Figure 2. Absorption spectra for pre-production samples of Gd-LS immediately after preparation (solid line) and 1172 days later (dashed line).

The Gd-LS fabricated during the test runs has been monitored for more than 3 years. The consistency of the optical transparency over time (see Figure 2) demonstrated excellent long-term stability. The production system and the synthesis procedures were thus qualified, together with other characterizations.

**2. Gd-LS production**

The main raw materials of the Daya Bay 0.1% Gd-LS and LS are 388 tons of LAB (specialized, better than detergent grade, Jinling Petrochemical Ltd.), 1.1 tons of gadolinium chloride ($GdCl_3$ $xH_2O$, 99.99%, Stanford Materials), 1.2 tons of TMHA (99% GC grade, CHEMOS GmbH), 1.4 tons of PPO (JINR), and 7 kilograms of bis-MSB (scintillation grade, RPI). Small amounts of hydrogen chloride and ammonia water used for pH adjustment are ultrapure reagents from Beijing Institute of Chemical Reagents.

All parts of the production system were cleaned with detergent, rinsed with deionized water, and dried with $N_2$ gas before packing. All pipes, valves, and tanks were again flushed with deionized water and dried with $N_2$ before installation at the Daya Bay LS Hall.

Gd-LS is fabricated in four main steps. Solutions of purified $GdCl_3$ and TMHA are prepared in separate acrylic tanks and transferred by metering pumps simultaneously to a 1000-L reaction tank where the Gd-TMHA solid precipitates. The solid is washed with pure water and the damp material is then dissolved directly into LAB to give a [Gd] concentration of 0.5% by mass. The fresh Gd-LAB is transferred to a 1-ton clearance tank for the separation of water droplets remaining after the synthesis process. In the meantime, a concentrated batch of fluor and shifter (PPO and bis-MSB at 10× higher than target concentration) in LAB is prepared. Both the 0.5% Gd-LAB (after water separation and removal) and the concentrated PPO-bis-MSB-LAB solutions are then transferred to a 5000-L dilution tank for final mixing. Additional LAB is added to make a 4-ton batch of 0.1% Gd-LS. Filtration (Teflon column at 0.5-micron) is applied at every stage of liquid transfer. An $N_2$ gas purge during the final mixing in the 5000-L dilution tank and periodically in the 40-ton storage tanks prevents oxygen and radon intrusions. Details are described in the following subsections.

2.1 Purification

Purification of raw material is crucial to the quality of the Gd-LS and LS. Two types of impurities are of most concern: one is colored contaminants that have significant impacts on optical transparency and are potential threats to the chemical stability of scintillator; the other is radioactive daughters from the naturally occurring uranium (U) and thorium (Th) decay chains. The radioactivity must be reduced to the sub-ppb level for the Daya Bay experiment. All purifications must be completed before the starting of synthesis.

It should be noted that since LAB is the main component (99%) of the 0.1% Gd-LS, it has the largest impact on performance of the scintillator. The manufacturer optimized the production flow, changed to new catalyst, and used temporary PTFE transfer pipes so that a high quality batch of LAB was produced within 3 days in Oct. 2009. Nearly

400 tons of LAB were delivered to the Daya Bay LS Hall and stored in the nylon bags in the concrete storage pools described above.

### 2.1.1 Purification of $GdCl_3 \cdot xH_2O$

Raw $GdCl_3 \cdot xH_2O$ often contains radioactive impurities of U, Th, and their decay daughters from the original mining ores. These impurities can produce events in an AD that mimic those from an antineutrino. In particular, the alpha particles emitted by the U/Th decay chains can produce α-n correlations through $^{13}C(\alpha,n)^{16}O$ interactions. With detector Monte Carlo simulations, the U/Th impurities in the raw $GdCl_3 \cdot xH_2O$ (assuming the nominal formula $GdCl_3 \cdot 6H_2O$) are required to be less than 1 ppb. The U/Th radioactivity in the raw $GdCl_3 \cdot xH_2O$ was measured to be in the range of 1 - 3 ppb (batch varied), higher than the requirement of 1 ppb.

Other colored impurities, such as iron (Fe) and cobalt (Co), must be removed to minimize their impacts on optical transparency and chemical stability. A bench-top experiment was performed at IHEP by adding an Fe spike to $GdCl_3 \cdot xH_2O$ before synthesis to determine its effect on the scintillator. The Gd-TMHA solid synthesized with Fe was then dissolved in LAB at concentrations of 0.1% and 0.004% for Gd and Fe respectively. The Fe-contaminated Gd-LS was monitored by UV-Vis spectrometer for 3 months. Another 0.1% Gd-LAB without Fe-addition was also prepared as reference for comparison. Figure 3 shows the difference in transparency of 0.1% Gd-LAB with or without Fe. The presence of iron not only increases absorbance in the region from 340 nm to 600 nm, but also affects the chemical stability of Gd-LAB as a function of time.

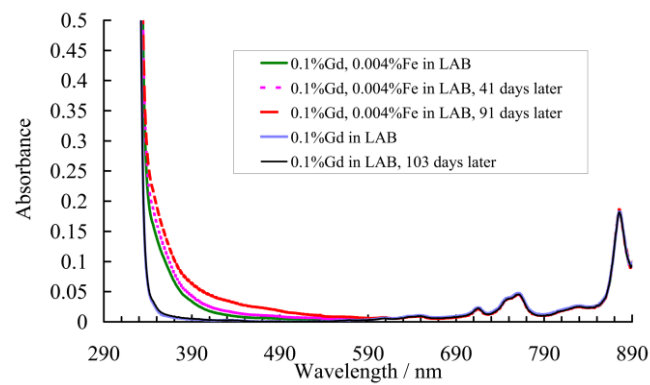

Figure 3. Comparison of absorption spectra of 0.1% Gd-LAB with or without Fe at various times after preparation.

A self-scavenging purification using pH adjustment and fine-filtration of $GdCl_3 \cdot xH_2O$ dissolved in water was developed for the production of the Daya Bay scintillator [14]. This procedure removes U, Th, and other colored impurities effectively. Nearly 1 kg of $GdCl_3 \cdot xH_2O$ was purified at BNL and assessed at the LBNL underground facility. The results indicate that a U/Th level below 1 ppb was achieved. The optical transparency of self-scavenged $GdCl_3 \cdot xH_2O$ was also measured by the 10-cm path-length UV-Vis spectrometer. Absorption spectra of Gd aqueous solutions before and after purification are presented in Figure 4. Data show that the self-scavenging purification not only removes the radioactive isotopes, but also improves the optical transparency significantly. No trace of Fe can be detected after purification. The long-lived radium in the U/Th chain cannot be removed by the pH adjustment method but won't enter Gd-LS since they are unlikely to complex with TMHA.

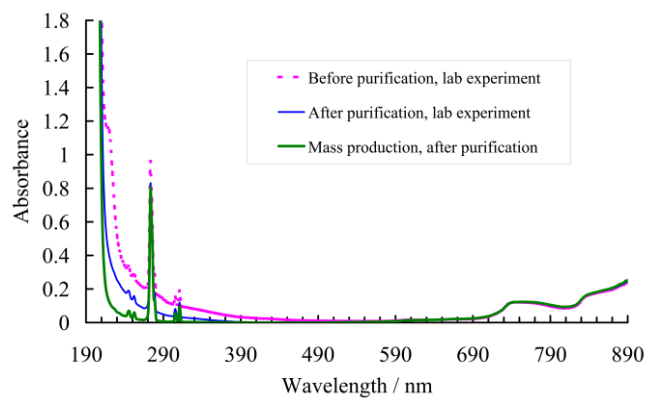

Figure 4. Absorption spectra of Gd aqueous solutions before and after self-scavenging purification.

2.1.2 Purification of TMHA

TMHA is an organic weak acid with a boiling point of ~230$^{o}$C under ambient pressure. A sum of 1.2 tons of THMA was purified by thin-film vacuum distillation operating in a batch mode of 20 kg per day. The UV spectra of TMHA before and after purification are compared in Figure 5. The results show that this vacuum distillation improves the optical transparency of TMHA by a factor of 2 in the PMT sensitive region.

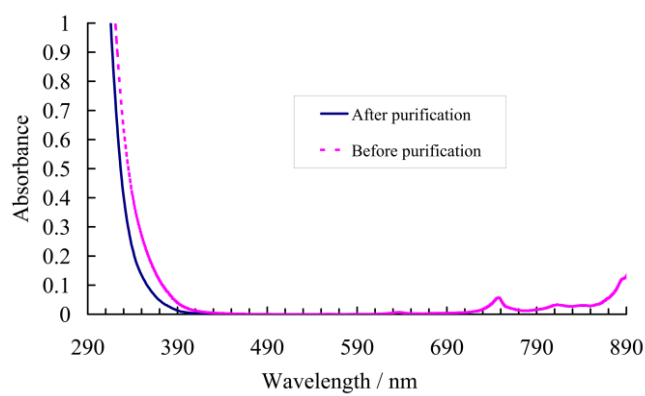

Figure 5. Absorption spectra for TMHA before and after purification.

2.1.3 Purification of PPO

The concentration of PPO, 3g/L, is the highest of the solutes in the Gd-LS. Accordingly radioactive and colored impurities in it are of main concern for the quality of Gd-LS. The required amount of PPO was provided by collaborators at the Joint Institute for Nuclear Research (JINR), Russia. Additional purifications including filtration after melting, distillation, and recrystallization, were contracted to an analytical laboratory (Huashuo Technology Co., Ltd.) in Wuhan, China. Purified samples must pass Daya Bay quality assurance checks before shipment. A total of 1.4 tons of PPO has been purified in different lots over three months. The results show that purified PPO met our QA requirements (see Figure 6).

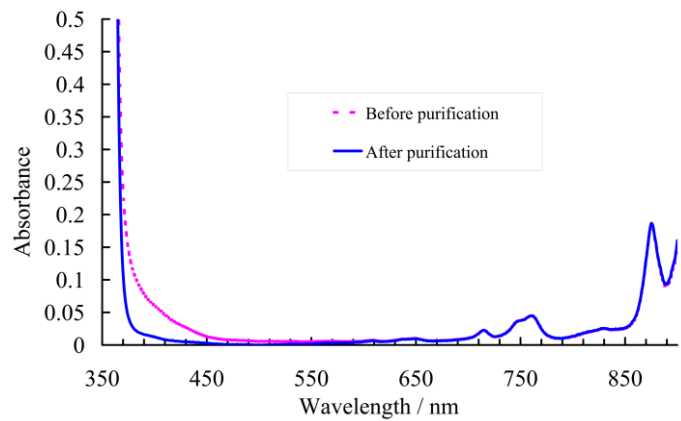

Figure 6. Absorption spectra of LAB containing 10 g/L PPO before and after purification.

2.2 Water

A water purification system was built for the Gd-LS production in the tunnel outside the LS Hall. When compared with 18.2-MΩ·cm ultrapure water from a bench-top Millipore Ultrapure System, the Daya Bay purified water (~15 MΩ·cm) showed slightly higher absorbance in the visible region (see Figure 7). No detectable metallic impurities were identified in the Daya Bay purified water.

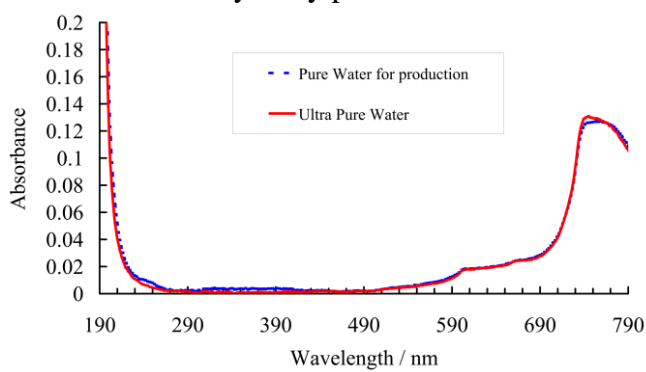

Figure 7. Comparison of absorbance spectrum of purified water for Gd-LS production (dashed curve) with that for ultrapure water from a Millipore system (solid curve).

2.3 Synthesis of Gd-TMHA and QA/QC

The Gd-LS production scheme consists of the formation of a moist Gd-TMHA solid, followed by its dissolution in LAB to make a concentrated batch of 0.5% Gd-LAB, and then by a 5× dilution to get 0.1% Gd-LS. The Gd-$H_2O$ and TMHA-$NH_4OH$ solutions were transferred at the ratio of 1:1 by two metering pumps for the synthesis of Gd-TMHA solid in the 1000-L reaction tank. The pH adjustment of TMHA neutralization is one of the key parameters for synthesis and must be controlled precisely. The averaged deviation of the pH of THMA-$NH_4OH$ solution is controlled at 1% variation over 50 batches of production.

2.4 Dissolution of Gd-TMHA and QA/QC

Each batch of Gd-TMHA solid was washed three times with purified water to remove traces of residual reactants from the solid before dissolution. Most water from the synthesis process and washings was drained out through the bottom filter of the 1000-L reaction tank. The Gd-TMHA solid was left in the reaction tank overnight to further drain off as much residual water as possible. The amount of LAB required to make 0.5% Gd by mass was then added to the tank to dissolve the Gd-TMHA solid. Freshly dissolved Gd-LAB often appears opaque initially due to tiny suspended water droplets in the organic phase. It was transferred through a nylon filter, which helps the demulsification, to other 1-ton clearance tanks for further water separation by gravity. After ~1 day of standing in the tank, depending on the residual water content of the Gd-TMHA solid, agitation speed, and dissolution time, the Gd-LAB solution turned clear as being observed in the test runs described above. The concentration of [Gd] averaged over 50 batches is 0.507±0.025% by mass.

2.5 Fluors-Shifter-LAB and QA/QC

Instead of adding PPO and bis-MSB powders directly to the Gd-LAB solution, a concentrated solution of PPO (30 g/L) and bis-MSB (150 mg/L) in LAB was prepared before the final mixing of the 0.1% Gd-LS. This simplifies the process and ensures a consistent PPO-bis-MSB content in each 0.1% Gd-LS batch.

2.6 Dilution of the 0.1% Gd-LS and QA/QC

The final mixing of the 0.1% Gd-LS was completed in the 5000-L dilution tank as described in section 2. The concentrated Gd-LAB (from 3.4) and PPO-bis-MSB-LAB (from 3.5) solutions were transferred into this dilution tank with additional LAB to match the final concentration of [Gd] at 0.1%. A nitrogen purge was introduced during the dilution process to remove oxygen, a known quencher which can reduce the light yield of the scintillator by as much as ~11% [15].

Gadolinium concentration and optical transparency of each batch of the 0.1% Gd-LS were characterized by XRF and a 10-cm path-length UV spectrometer, respectively, before piping into the storage tanks installed next to the production station. The [Gd] concentration over 50 batches of Gd-LS were measured to be within 5% of 0.100%, while the measurement uncertainty of XRF is about 2%. The optical property of each Gd-LS batch was measured by two UV-Vis spectrometers, Shimadzu UV-2550 and UV-1800, for crosscheck. The data between two spectrometers agree within 1% at

0.0015±0.0005 at 430 nm over 50 batches. Small variations in [Gd] and optical properties are due to slightly different synthesis conditions from batch to batch. However the difference is averaged out once the batches are transferred into the 40-ton storage tanks and circulated during storage. When filling into an AD of 20-ton target mass, 4 ton Gd-LS from each of the five storage tanks was taken to ensure the [Gd] concentration is the same for all ADs.

## 3. Characterization and Long-Term Monitoring of Gd-LS

To prevent intrusions of oxygen and radon, a clean $N_2$ buffer was applied to the five 40-ton storage tanks with a slight positive pressure. Each tank has a built-in circulation system to mix the Gd-LS with a flow rate of 4-ton per hour for more than 50 hours. This corresponds to five volume exchanges to assure homogenous circulation of the Gd-LS in each storage tank. The quality of the Gd-LS from the storage tanks has been monitored constantly. The averaged [Gd] content is 0.103±0.002% by mass and has been stable over 500 days (see Figure 8).

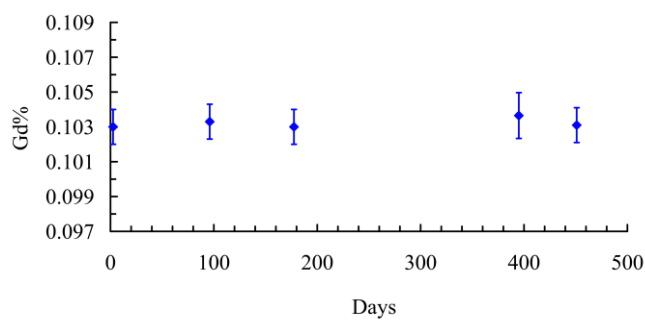

Figure 8. [Gd] content of Gd-LS as a function of time.

3.1 Optical Transparency

The transparency of the Gd-LS from each storage tank was monitored periodically for stability checks. Figure 9 shows that absorbance spectra are identical for the Gd-LS from the 5 storage tanks. Figure 10 presents the optical transparency of the Gd-LS at 430 nm as a function of time. The absorption value is stable at 0.0016±0.0002 over 500 days.

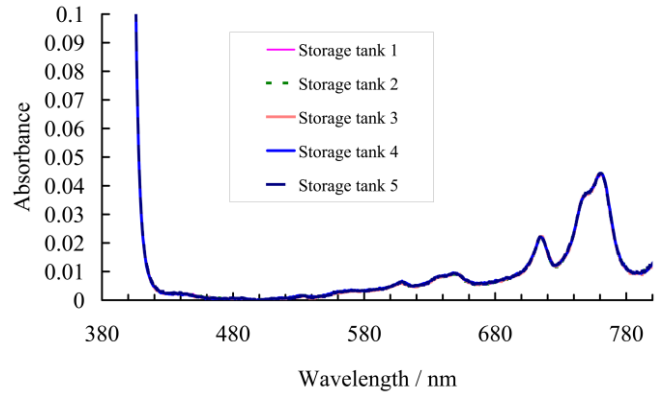

Figure 9. Comparison of absorbance spectra of Gd-LS samples from five 40-ton storage tanks at Daya Bay.

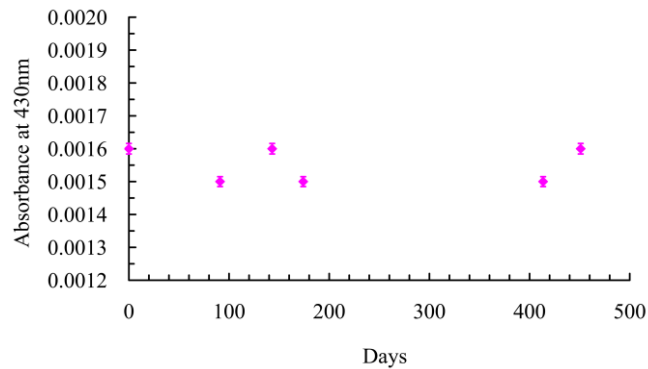

Figure 10. Optical absorbance at 430 nm of Gd-LS as a function of time.

3.2 Attenuation Length (*1/e*)

Attenuation lengths of the Gd-LS were measured by three long path-length systems: a 1-m semi-automatic apparatus on-site at Daya Bay by LBNL, a 2-m dual-beam system off-site at BNL [16] and a 1-m system off-site at IHEP [17]. The attenuation length of the Gd-LS in the storage tanks was measured to be ~15 meters at 430 nm with the above three apparatus.

The ADs in place are calibrated weekly with sources and LED. The attenuation length of the Gd-LS can also be monitored in-situ with the calibration data. A slow degradation, ~1.3% per year, of the energy scale in terms of photoelectrons per MeV was observed [18]. The ADs are cylindrical structures with 192 8-inch photomultiplier tubes (PMTs) mounted along the circumference of the detector [3]. By comparing the PMT charge patterns for sources at different vertical positions, attenuation length decrease of the Gd-LS and/or LS is identified as the reason of this degradation. Current data is not enough to identify if it is due to contamination from detector materials, copolymerization of the Gd-complex, oxidation of the liquid scintillator, or any other reasons. The [Gd] concentration monitored by the neutron capture time keeps very stable. The slow degradation of the energy scale will not introduce obvious uncertainties for the neutrino oscillation studies during the whole lifetime of the Daya

Bay experiment.

## 3.3 Emission Spectra and Light-Yield

The fluor and shifter for the Daya Bay scintillator are PPO at 3 g/L and bis-MSB at 15 mg/L. The emission spectrum (by PTI fluorescence spectrometry) from the Gd-LS excited at 260 nm is shown in Figure 11, agreeing well with the bis-MSB emission spectrum. The light yield of this LAB-based liquid scintillator is high, ~50% of that of anthracene for both Gd-LS and LS, as measured off-site at both BNL and IHEP.

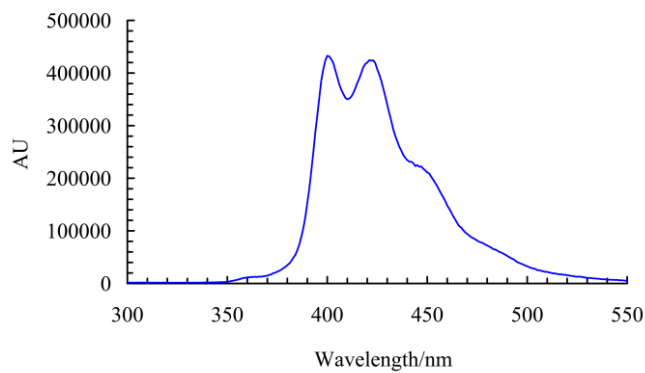

Figure 11. The emission spectrum of Gd-LS when excited at 260 nm.

## 3.4 Backgrounds from Gd-LS

The rate of >0.7 MeV singles was measured to be ~60 Hz in all ADs. The contribution by the radioactive impurities in the Gd-LS has been measured to be ~4.5 Hz by applying a tight volume cut then scaling to the total Gd-LS volume. The $^{13}C(\alpha,n)^{16}O$ background rate was estimated to be 0.04-0.08 per day for different ADs by measuring the alpha-decay rates in-situ and then using the MC calculation of neutron yield [18]. The low backgrounds measured in data demonstrate the successes of the raw material purifications and the cleanness control throughout the Gd-LS production, storage and filling processes.

## 4.5 Neutron capture time and Gd capture ratio

The relative difference of the fraction of neutrons captured by Gd in the different detectors is required to be <0.1% [1]. For each operational AD, the neutron capture time was measured using $^{241}$Am-$^{13}$C neutron source, IBD neutrons, and spallation neutrons. The relative difference between ADs was determined to be within 0.15 μs. This indicates that the Gd capture fraction variation among ADs is within 0.1% [3,18]. In addition, two temporary special calibrated neutron sources, $^{241}$Am-$^9$Be and $^{239}$Pu-$^{13}$C, were deployed in the first two ADs in August, 2012. High statistics of neutron capture events were collected. A side-by-side comparison of the directly measured Gd capture ratios was performed between two ADs. The relative difference was also determined to be less than 0.1%. The neutron capture time monitored periodically by various neutron sources and continuously by IBD and spallation neutrons confirms the stability of Gd concentration.

## 4. Conclusion and Discussion

The Daya Bay liquid scintillator has been prepared consistently with precise quality control. Purification of the principal starting materials was applied to each batch of production. One hundred and eighty-five tons of 0.1% Gd-LS and two hundred tons of LS were successfully produced on schedule from Oct. 2010 to March 2011. The Gd-LS liquid properties are well-characterized and essentially constant to within 1% between different storage tanks before filling the ADs. Good Gd-LS attenuation length and high light-yield meet the Daya Bay quality assurance requirements. Equal portion of scintillator (4 tons) from each storage tank was taken for filling each 20-ton AD to ensure identical functions of all eight detectors. The variation of Gd capture ratio among ADs was determined to meet the 0.1% design specification by measuring the neutron capture time. The long-term stability monitoring program for Daya Bay scintillator, started three months after production, has been ongoing for more than 2 years and will be continuing along-side the Daya Bay data-taking. The long-term chemical stability of the Gd-LS has surpassed the previous reactor antineutrino experiments and led to a successful discovery of non-zero $\theta_{13}$ in 2012 [2].

**Acknowledgements:**

We would like to express our gratitude to the technical staffs of IHEP for excellent support in production, to Raymond Kwok and his colleagues at CUHK for sample arrangement and shipment for long-term stability monitoring, and to Jinchang Liu, Chengju Lin, WeiliZhong, and J. Pedro Ochoa from IHEP and LBNL for interesting discussions and for help in the attenuation length assessments. This work was partially supported by the Ministry of Science and Technology of China (Grant No. 2013CB834300), National Natural Science Foundation of China (Grant No. 11005117), the U.S. Department of Energy, Office of High Energy Physics and Office of Nuclear Physics (under Contract Nos. DE-AC02-98CH10886 and DE-AC02-05CH11231).